\documentstyle[12pt]{article}
\begin{document}

\title{
The Isospin Distribution of Fragments in Reactions 
$^{96}Ru+^{96}Ru$, $^{96}Ru+^{96}Zr$, $^{96}Zr+^{96}Ru$, and $^{96}Zr+^{96}Zr$ at
Beam Energy 400 AMeV  \footnote{Supported by
National Natural Science Foundation of China under No. 19975073 and 10175093, National Science Foundation 
of Nuclear Industry under No. Y7100A0101 and Major State 
Basic Research Development Program in China under contract No. G20000774}}
\date{}
\maketitle

\begin{center}
\author{\normalsize Qingfeng Li$^{1)}$, Zhuxia Li$^{1,2,3)}$\\
\small 1) China Institute of Atomic Energy, P. O. Box 275 (18), 
Beijing 102413, P. R. China\\
2) Institute of Theoretical Physics, 
Academia Sinica, P. O. Box 2735, Beijing 100080, P. R. China\\
3) Center of Theoretical Nuclear Physics, National
Laboratory of Lanzhou Heavy Ion Accelerator, Lanzhou 730000, P. R. China
}
\end{center}

\begin{abstract}
   The isospin distribution of particles and fragments in collisions $^{96}Ru+^{96}Ru$, 
 $^{96}Ru+^{96}Zr$, $^{96}Zr+^{96}Ru$, and $ ^{96}Zr+^{96}Zr$ at beam
 energy 400 AMeV is studied with isospin dependent QMD model. 
We find that the rapidity distribution of differential neutron-proton counting in 
neutron rich nucleus-nucleus collisions at intermediate energies is sensitive to 
the isospin dependent part of nuclear potential. The study of the N/Z ratio of nucleons, light charged 
particles (LCP) and intermediate mass fragments (IMF) shows that the isospin dependent part of nuclear 
potential 
drives IMF to be more isospin symmetric and emitted nucleons to be more neutron rich. 
From the study of the time evolution of the isospin distribution in emitted nucleons, 
LCP and IMF we find that neutrons diffuse much faster than protons at beginning  
and the final isospin distribution is a result
of dynamical balance of  symmetry potential and Coulomb force  under the  charge conservation. 
\end{abstract}



Nuclear multifragmentation has been received more and more attention in both 
theoretical and experimental studies since people believe that it carries abundant
information of nuclear dynamics and the properties of nuclear matter, 
especially, the possible association with liquid-gas phase transition.  
Following the establishment of radioactive beam facilities at many laboratories,
 the study of nuclear multifragmentation for isospin asymmetric nuclear system 
becomes possible. For isospin asymmetric systems, a description of bi-component 
system should be introduced while a bi-component system will manifest a richer 
thermodynamical behavior. The new features of phase separation process for asymmetric nuclear 
system have been discussed earlier by Barranco and Buchler \cite{Ba80} with phenomenological equation of 
state and later by M\"uller and Serot \cite{Se95} with relativistic mean field model.    
It was pointed out in \cite{Ba80,Se95} that  the spinodal is defined by the chemical 
instability  rather than  by mechanical instability for isospin asymmetric systems
, which allows gas and liquid to have different concentration of protons and neutrons. 
The dynamical properties of asymmetric nuclear matter were investigated in \cite{Ba98,Ba01,Ku96} and 
it was predicted that the spinodal decomposition is accompanied by a collective diffusion of protons from low- to high-density region. 
Experimentally, it was found that the N/Z ratio of gas significantly exceeds that of 
liquid in the multifragmentation process in heavy ion collisions \cite{Xu00,Ts01,Ve00}.  
It was further showed that the properties of liquid-gas phase transition 
depend on the value of the symmetry energy coefficient\cite{Se95,Ku96}. 
As is well known, the symmetry energy term 
of EOS ( both the density dependence of symmetry potential energy and the 
value of symmetry energy coefficient) has large uncertainties. For example, the theoretically predicted value of
symmetry energy coefficient  is about $27 - 38 MeV$ by non-relativistic Hartree-Fock 
approach \cite{Pe91},  $35 - 40 MeV$ by relativistic mean field approach \cite{Ma81,Ru88,Fu95}, 
$31 MeV$ by Brueckner-Hartree-Fock theory (BHF) \cite{Lee98} and $28.7MeV$ by  extended BHF theory \cite{Zu99}. 
Furthermore, recent study has showed that the 
symmetry energy coefficient increases as the isospin asymmetry increases and the increasing slope 
is rather different for different version of Skyrme force \cite{Br01}.  
Therefore it might be interesting 
to study the sensitivity of various observables of nuclear multifragmentation
 to symmetry energy coefficient. 
We will first   
investigate the dependence of the various observables in multifragmentation 
on the symmetry energy coefficient and try to find the sensitive observables to 
the symmetry energy coefficient. 
Then the effects of the isospin dependent part of nuclear potential on
the isospin distribution of  emitted nucleons, LCP and IMF in multifragmentation are  studied.   
 In order to study the dependence of the isospin distribution on the 
isospin asymmetry of the system we perform a set of calculations for the mixing reactions 
of four mass 96+96 systems Ru+Ru, Zr+Zr, Ru+Zr, and Zr+Ru with corresponding N/Z ratios of 
1.18, 1.4, 1.28, 1.28.  It is noticed that $^{96}Ru$ is of 6 neutron deficiency 
and $ ^{96}Zr$ is of 6 neutron excess compared with the most stable isotopes 
$^{102}Ru$ and $^{90}Zr$ .\\ 

The isospin dependent quantum
molecular dynamics (IQMD) model \cite{Ai91,Har89} is used in the calculations. 
In this model, each nucleon is represented by a Gaussian wave packet centered at 
 ${\bf r_{i}}$ and ${\bf p_{i}}$. 
The time evolution of ${\bf r_{i}}$ and ${\bf p_{i}}$ is
governed by Hamiltonian equation of motion:
\begin{equation}  \label{eq1}
 \dot{{\bf r}}_{i}=\frac{\partial H}{\partial {\bf p}_{i}}, 
 \dot{{\bf p}}_{i}=-\frac{\partial H}{\partial {\bf r}_{i}}.
\end{equation}

The Hamiltonian $H$ consists of the kinetic energy and effective
interaction potential energy,
\begin{equation}  \label{eq2}
H=T+U.
\end{equation}
The effective interaction potential energy includes the nuclear
local interaction potential energy and Coulomb interaction
potential energy,
\begin{equation}  \label{eq3}
U=U_{loc}+U_{Coul}.
\end{equation}
The local interaction potential energy can be obtained by   
\begin{equation}  \label{4}
U_{loc}=\int V_{loc}d^{3}r,
\end{equation}
where $V_{loc}$ is the potential energy density which can be obtained from Skyrme type interaction and reads as: 
\begin{equation}  \label{5}
V_{loc}=\frac{\alpha }{2}\frac{\rho ({\bf r})^{2}}{\rho _{0}}+\frac{\beta }{3}%
\frac{\rho ({\bf r})^{3}}{\rho
_{0}^{2}}+\frac{C_{s}}{2}\frac{(\rho _{p}({\bf r})-\rho _{n}({\bf
r}))^{2}}{\rho _{0}}.
\end{equation}

The third term in the right side of equation (5) is the symmetry potential energy density, in which a linear density dependence of 
for the symmetry potential energy is adopted . In general,  $V_{sym}$ can be expressed as: 
\begin{equation} \label{eq6}
V_{sym}=\frac{C_{S}}{2}\frac{(\rho _{n}-\rho _{p})^{2}}{\rho^2}\rho F(u). 
\end{equation}
 Here $F(u)$ gives the density dependence of symmetry potential energy and $u=\rho/\rho_0$
. Phenomenologically, Prakash, Ainsworth, and Lattimer proposed the forms of F(u) as $u$, $u^{1/2}$, and $2u^2/(1+u)$ \cite{Pr88},respectively. 
For  the form of  $u^{1/2}$ or $2u^2/(1+u)$, we can hardly  get the analytical expression of $U_{sym}$ from expression (4).
However, it has been shown  in many microscopic studies  
that the linear density dependence of symmetry energy was almost valid at not  very far from  normal densities. 
 (see, for example, \cite{Lee98,Zu99}). Thus  in this work a linear density dependence 
of symmetry potential energy, $F(u)=u$, is  still adopted for the convenience of  performing
IQMD calculations. Of course  the knowledge of  the density dependence of  the symmetry potential energy is 
very much concerned, which will be studied in the future.
 The symmetry potential energy term in equation (5)  can be re-written as 
\begin{equation}
V_{sym}(\rho, \delta )=\frac{C_S}{2} (\frac{\rho}{\rho_{0}}) \delta^2 \rho,  \label{eq8}
\end{equation}
where

\begin{equation}
\delta =\frac{\rho_n-\rho_p}{\rho_n+\rho_p}.  \label{eq9}
\end{equation}
The corresponding symmetry energy coefficient is
\begin{equation}
a_{sym}=\frac{3}{5}(2^{2/3}-1)\epsilon_{F}^{0} + \frac{C_S}{2}
       \simeq\frac{\epsilon_{F}^{0}}{3} + \frac{C_S}{2}.  \label{eq10}
\end{equation}  
Here $\epsilon_{F}^{0}$ is the Fermi energy of the symmetric system at $\rho=\rho_{0}$ and 
taken to be 38 MeV.   
The effect of the isospin dependent part of nuclear potential on the isospin distribution in multifragmentation 
is tested by taking different values of $C_S$ in the calculations. In this work we take $C_S$ equals $0$ MeV, $27 MeV$, $35 MeV$ and $50 MeV$, 
corresponding to the values of symmetry energy coefficient $a_{sym}$: $0$ MeV,$27 MeV$, $31 MeV$ and $38 MeV$,respectively.\\
 
The experimental isospin dependent binary elastic scattering cross section 
is used \cite{Chen77} in  calculations. In \cite{Chen77} it  was shown that up to hundreds MeV the elastic proton-neutron cross section
is about $2 - 3$ times larger than that of proton-proton (neutron-neutron)'s.\\
 
Concerning the treatment of the Pauli blocking, we firstly distinguish protons and
neutrons, and then we use the following two criteria:

\begin{equation}
\frac{4\pi}{3} r_{ij}^{3}\cdot \frac{4\pi}{3}p_{ij}^{3}\geq
\frac{h^{3}}{4},  \label{eq11}
\end{equation}
and

\begin{equation}
P_{block}=1-(1-f_{i})(1-f_{j}),  \label{eq12}
\end{equation}
where $f_{i}$ is the distribution function in phase space for particle $i$
and reads as
\begin{equation}
f_i(\stackrel{\rightarrow }{r},\stackrel{\rightarrow }{p},t)=\frac {1}{\pi
\hbar^{3}}\exp (-(\stackrel{\rightarrow }{r}-\stackrel{\rightarrow }{r_i}
(t))^{2}/2L^{2})\exp (-(\stackrel{\rightarrow }{p}-\stackrel{\rightarrow }
{p_i} (t))^{2}2L^{2}/\hbar^{2}),  \label{eq13}
\end{equation}
where L is a parameter which represents the spatial spread of wave packet, 
and $\stackrel{\rightarrow }{r_i}(t)$ and $\stackrel{\rightarrow }{p_i}(t)$
denote the center of the wave packet of particle $i$ in coordinate and momentum space
respectively. The first condition(expression (10))  gives the criterion for the uncertainty relation of 
the centroids of Gaussion
wave packets of two particles. The second one is the probability of the 
Pauli blocking effect for the scattering of two particles, which is
especially useful for collisions of heavy nuclei. In this paper, we construct 
clusters in terms of the conventional coalescence model \cite{Kruse85}, 
in which particles with relative momenta smaller than $P_0$ and relative distances 
smaller than $R_0$ are considered to belong to one cluster.
In this work  $R_0$ and $P_0$ are taken to be 3.5 fm and 300 MeV/c,respectively, following Ref.\ \cite{Zhang99}. 
In addition, only the cluster with reasonable proton number Z and neutron number N are selected in 
order to get rid of nonphysical clusters.

The soft EOS ($K=200 MeV$) is used in the calculations, and the
corresponding main parameters are listed in Table 1.

First of all, in order to check our IQMD model, we calculate the rapidity
distribution of protons to make comparison with experimental data 
of \cite{FOPI00}.
Fig.\ 1 shows the calculated rapidity distribution of protons (including 
deuterons for comparing  with experiments) for four colliding systems: Zr+Zr (solid line), Ru+Ru (dashed line), 
Zr+Ru (dotted line) and Ru+Zr (dash-dotted line) at $E=400 AMeV$ and $b=0 fm$
,in which $C_S=35MeV$  is used and the angle selection 
($10^{0}<\theta_{lab}<28^{0}$ and $34^{0}<\theta_{lab}<145^{0}$) is taken which is the same  with that of   
the experimental data \cite{FOPI00}.
 The experimental data are also given in the figure (solid circles and open circles reflected
from experimental data assuming a backward/forward symmetry). 
This figure shows that our calculation results can reproduce the
experimental data quite well, only at the projectile and target region, the
calculated results deviate from experimental data, which is because the
calculated results are for head on collisions while the experimental data
are for central collisions. 
It is also shown in the figure that the rapidity distribution of protons for Ru+Ru is
higher than that for Zr+Zr at whole rapidity region because of the difference of 
their initial proton number. The rapidity distribution for asymmetric reactions 
Zr+Ru and Ru+Zr is between that of Zr+Zr and Ru+Ru.
It is checked  that the results do not
change much when $C_S$ is taken to be from $27 MeV$ to $ 50 MeV$. 

\[
\fbox{Fig.\ 1} 
\]

Now let us investigate the sensitivity of the proton number counting to the value of 
$C_{S}$ because the proton number counting for the mixing reactions 
of four mass 96+96 systems Ru+Ru, Zr+Zr, Ru+Zr, Zr+Ru was experimentally
measured by FOPI/GSI recently \cite{Re99,FOPI00}.  
In \cite{FOPI00} a normalized proton number counting $R_{Z}$ is introduced:
\begin{equation}
R_{Z}=\frac{2*Z-Z^{Zr}-Z^{Ru}}{Z^{Zr}-Z^{Ru}}.  \label{eq14}
\end{equation}
By definition $R_{Z}=1$ for Zr+Zr, $R_{Z}=-1$ for Ru+Ru. 
Fig.\ 2 shows the rapidity distribution of $R_Z$ calculated with $C_S = 0MeV$, $27MeV$, $35MeV$ and $50MeV$. 
And we can see that $R_Z$  
can explore the non-equilibrium effect in nucleus-nucleus colliding process 
successfully. However the results with $C_S $= $0 MeV$, $27 MeV$, $35 MeV$ and $50 MeV$, respectively, are 
indistinguishable. It means that the normalized proton number counting $R_{Z}$ 
is not sensitive to the value of symmetry energy coefficient, which may be understood from 
the definition of $R_Z$ where the effect of changing $C_S$ 
on $R_Z$ is significantly reduced. 
So we have to seek the other observables which are sensitive to 
$C_{S}$.

\[
\fbox{Fig.\ 2} 
\]

Here we introduce the normalized differential neutron-proton counting 
which is defined as:
 
\begin{equation}
N_{np}=\frac{<n>-<p>}{<n>+<p>}.  \label{eq15}
\end{equation}
Where $<n>$ and $<p>$ is the average number of emitted neutrons and  protons, respectively. 
The relation between $N_{np}$ and average N/Z ratio of emitted nucleons is 
\begin{equation}
N/Z= \frac{1+N_{np}}{1-N_{np}}.  \label{eq16}
\end{equation}

The rapidity distribution of normalized differential neutron-proton counting has been shown in  Figs.\ 3
with different $C_S$ of  $0 MeV$, $27 MeV$, $35 MeV$, and $50 MeV$, in which a), b), c), d) 
corresponds to reactions Zr+Zr, Ru+Ru, Zr+Ru, and Ru+Zr, respectively.
 The dependence of $N_{np}$ on $C_{S}$ is very pronounced at 
the projectile and target rapidity region. The general tendency is that $N_{np}$ 
increases with $C_{S}$. 
When the isospin dependent part of nuclear potential is switched off ($C_{S}=0$), 
the rapidity distribution of $N_{np}$ becomes flat for symmetric reactions 
Zr+Zr and Ru+Ru while for asymmetric reactions 
Zr+Ru and Ru+Zr, it becomes  an inclined line from Zr side to Ru side, which 
indicates the non-equilibrium effect in  reactions. 
And the N/Z ratio of emitted nucleons (by using Eq.\ 13) at central rapidity ($Y=0$)
is almost equal to the initial N/Z ratio. As soon as the isospin dependent part of nuclear potential 
is switched on 
 $N_{np}$ increases with $C_{S}$ for off central rapidity cases. And 
one can find from 
Fig.\ 3 a) and Fig.\ 3 b) that at projectile and target rapidity region 
the difference in $N_{np}$ calculated with  different $C_{S}$ is more pronounced for Zr+Zr  
than that for Ru+Ru.
It  indicates that the sensitivity to $C_{S}$ increases 
as the system becomes more neutron rich, which can be understood from the expression of  the symmetry potential energy term(see Eq.(7)).  
Therefore the rapidity distribution of normalized differential neutron-proton counting
in neutron rich nuclear collisions at intermediate energies can be considered 
to be a good candidate of sensitive observables to the symmetry energy coefficient. 
    
\[
\fbox{Fig.\ 3 }  
\]

The N/Z ratio of particles and fragments is relevant to  
 the mechanism of multifragmentation for isospin asymmetric nuclear systems. 
So we further study the N/Z ratio of emitted nucleons, 
LCP (the cluster of Z=1 or 2) and IMF (the cluster of Z=3-16) as well as the average N/Z ratio in regardless of 
fragments or particles at projectile ($\left| Y-1\right| <0.5$), 
central ($\left| Y\right| <0.5$) and target ($\left| Y+1\right|
<0.5 $) rapidity regions, respectively. 
Fig.\ 4 shows the  calculated results  with different $C_{S}$ for reaction Zr+Zr.
The general feature of  the N/Z ratio of particles and fragments is: the largest one is  the N/Z ratio of nucleons,
then that of LCP, and smallest is that of IMF among them.
One can further find from the figure that  the symmetry  potential  
drives the N/Z ratio of LCP and IMF to approach to unit 
from sides of $N/Z > 1$ and $N/Z < 1$,  and simultaneously the N/Z ratio 
of nucleons is driven to the higher value than initial one, i.e., to be  more neutron rich.
The larger the $C_{S}$ is the stronger the effect is. 
It implies that the diffusion 
of neutrons increases faster than that of protons as $C_{S}$ increases. However, the total number of 
neutrons and protons has to be conserved and there is a balance of the numbers of neutrons and protons 
in nucleons, LCP and IMF at central, projectile and target rapidity region. 
Therefore the  N/Z ratios of  all products in multifragmentation is interdependent. 
If one looks at Fig.\ 4 more carefully one can find that at the central rapidity
region, the N/Z ratio of LCP at central rapidity region decreases 
while the N/Z ratio of nucleons increases as  
$C_{S}$ increases but the former one  is more sensitive to 
$C_{S}$ than the later one. On the other hand, 
the N/Z ratio of IMF increases towards unit when the isospin dependent term of nuclear potential is 
switched on.  But generally, the dependence of N/Z ratio of IMF on $C_{S}$ at projectile and 
target rapidity region is a little weaker than that of 
nucleons at projectile and target region and LCP at central rapidity region. 
Therefore we may also consider the N/Z ratio
of LCP at central rapidity region to be a relatively sensitive observable 
to $C_{S}$. 
The similar tendency of the sensitivity of N/Z ratio of emitted nucleons, LCP and IMF 
to $C_{S}$ is found for reaction Ru+Ru. 

\[
\fbox{Fig.\ 4} 
\]

It is convenient for a transport model to study the time evolution of isospin
distribution, from which we can also understand more deeply the dynamical
balance of isospin distribution in emitted nucleons, LCP and fragments. 
In fig.\ 5, we show the 
time evolution of the average N/Z ratio of all particles and fragments 
a), nucleons b), LCP c), and IMF d). The time ranges from $30fm/c$ to $100fm/c$. 
Here $30fm/c$ is 
the time when the density in a sphere of $1 fm^3$ at center drops to  $\rho_{0}$. 
In Fig.\ 5, the corresponding N/Z ratios at different time are given by different line types.  
From this figure,  we firstly see that at $30fm/c$ the N/Z ratio for nucleons and LCP is larger than 
unit, especially the former one. But for IMF it is much smaller than unit( it should be noticed that the number of  IMF at time  is small).  
It means that neutrons diffuse much faster than protons, which simultaneously results in  emitted IMF being neutron deficient . 
 As time goes on, the N/Z ratio 
of nucleons goes down which means that more protons are emitted during this period. 
After $50 fm/c$, the N/Z ratio of  emitted nucleons decreases very slowly.  The final  
N/Z ratio of emitted nucleons at central rapidity zone  
closes to the initial N/Z ratio of colliding systems but still higher than initial one. 
While at projectile and target rapidity region it is still much higher than initial N/Z ratio. 
And at the same time span, the N/Z ratio of  LCP
and IMF firstly goes up far away from unit before $50fm/c$ because of emitting of more protons during 
$30fm/c$ - $50fm/c$, and then goes downward to approach to unit. 
So from the above  discussion about  the time evolution of the isospin distribution in multifragmentation 
for heavy ion collisions at the energy studied we can draw two conclusions: 1) neutrons diffuse much faster than protons so the number of  
neutrons emitted much larger than that of protons at the beginning , and 2) the final isospin distribution 
is a result of dynamical balance of symmetry potential and Coulomb force under the total charge number conservation 
which leads  to the N/Z ratio of  IMF more close unit and emitted nucleons more neutron rich.    

\[
\fbox{Fig.\ 5} 
\]

In summary, in this work we have introduced a normalized differential neutron-proton 
counting for rich nuclei heavy ion in collisions  at intermediate energies as a
sensitive observable to probe the symmetry energy coefficient. 
We have also studied the dependence of the average N/Z ratio of nucleons, LCP and IMF
at central, projectile and target
rapidity region on  $C_S$ and we find that at the energy studied in this work
the general effect of isospin dependent part of nuclear potential is to drive  emitted 
nucleons to be more neutron rich and  IMF  to be more isospin symmetry. 
The stronger the isospin dependent part of nuclear potential is the stronger the effect is. 
It is also shown that the N/Z ratio of LCP at central rapidity region is relatively more sensitive 
to $C_S$ than the others. The time evolution of the isospin distribution in 
multifragmentation process shows that neutrons diffuse faster than protons  
and the final isospin distribution is a result of dynamical balance of symmetry potential and Coulomb 
force under the total charge number conservation.  

\begin{description}
\item[{\tt Table.\ 1}]  Parameters used in calculations
\end{description}

\begin{tabular}{|c|c|c|c|c|c|c|}
\hline
$\alpha (MeV)$ & $\beta (MeV)$ & $\gamma $ & $\rho _0 (fm^{-3})$ & $K (MeV)$ & 
$L (fm)$ & $C_{Yuk} (MeV)$ \\ \hline
$-356$ & $303$ & $7./6.$ & $0.168$ & $200$ & $2.1$ & $-5.5$ \\ \hline
\end{tabular}

\newpage
\begin{center}
{\bf CAPTIONS}
\end{center}
\begin{description}
\item[{\tt Fig.\ 1}]  The rapidity distribution of protons and deuterons for 
            $^{96}Ru+^{96}Ru$, $^{96}Ru+^{96}Zr$, $^{96}Zr+^{96}Ru$, 
            $^{96}Zr+^{96}Zr$ at $E=400 AMeV$, $b=0fm$. The experimental data of \cite{FOPI00} 
            are also given in the figure. 
\item[{\tt Fig.\ 2}]  The rapidity distribution of proton counting number $R_{z}$
 with different $C_{S}$ for the same reactions
            given in Fig.\ 1. 
\item[{\tt Fig.\ 3}] The rapidity distribution of differential neutron-proton counting 
            with different $C_{S}$ for a) $^{96}Zr+^{96}Zr$, b) $^{96}Ru+^{96}Ru$, 
           c) $^{96}Zr+^{96}Ru$, and d) $^{96}Ru+^{96}Zr$ at 400 AMeV, $b=0 fm$. 

\item[{\tt Fig.\ 4}] The average N/Z ratio of emitted nucleons, LCP and IMF
                at  projectile rapidity region, central rapidity region, 
                and target rapidity region with different $C_{S}$
            for $^{96}Zr+^{96}Zr$ at 400 AMeV, $b=0 fm$. 

\item[{\tt Fig.\ 5}] The time evolution of average N/Z ratio of emitted nucleons, LCP 
            and IMF at projectile rapidity region, center rapidity region, 
                 and target rapidity region
            for the same reactions as Fig.\ 4.
\end{description}
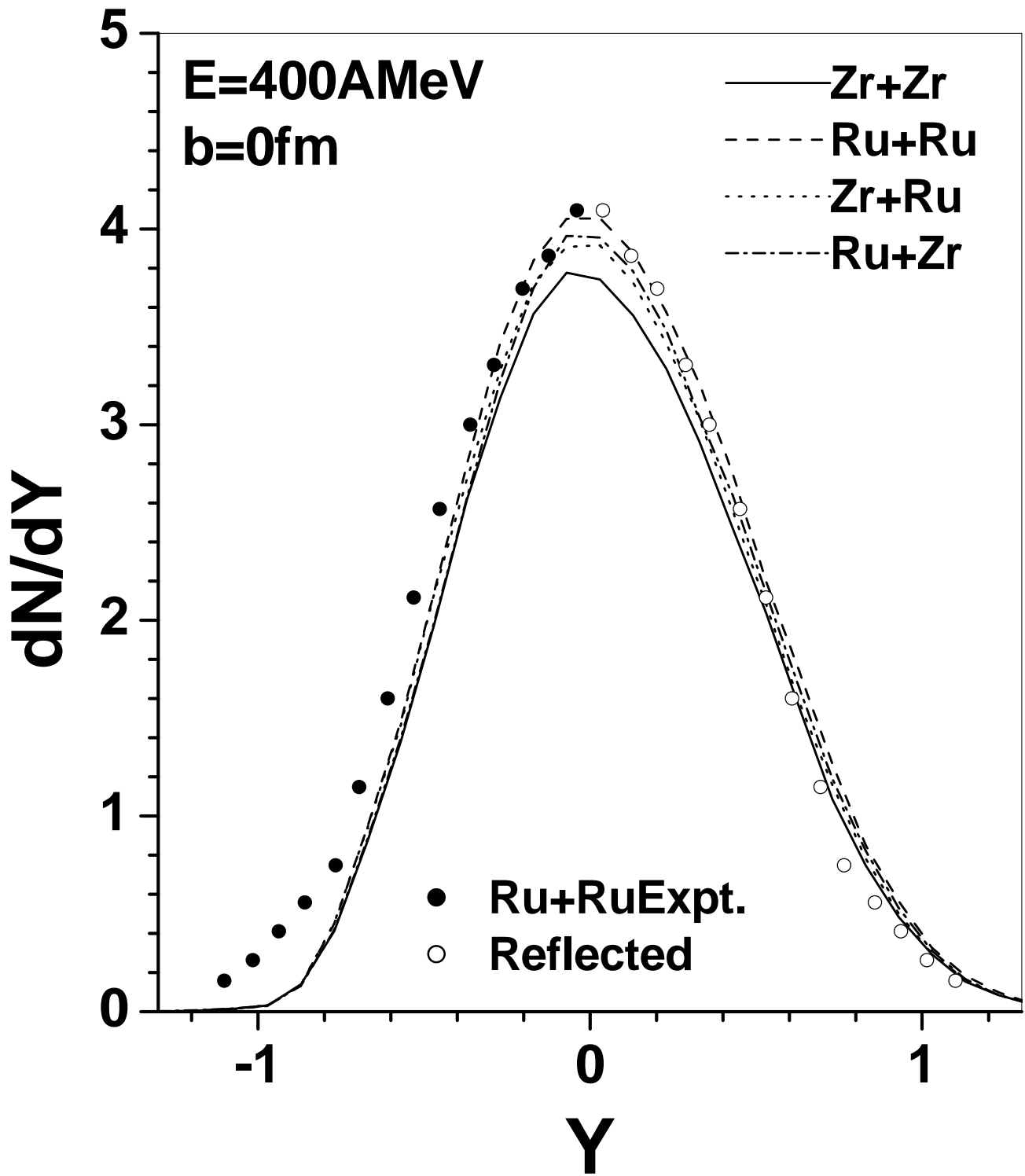
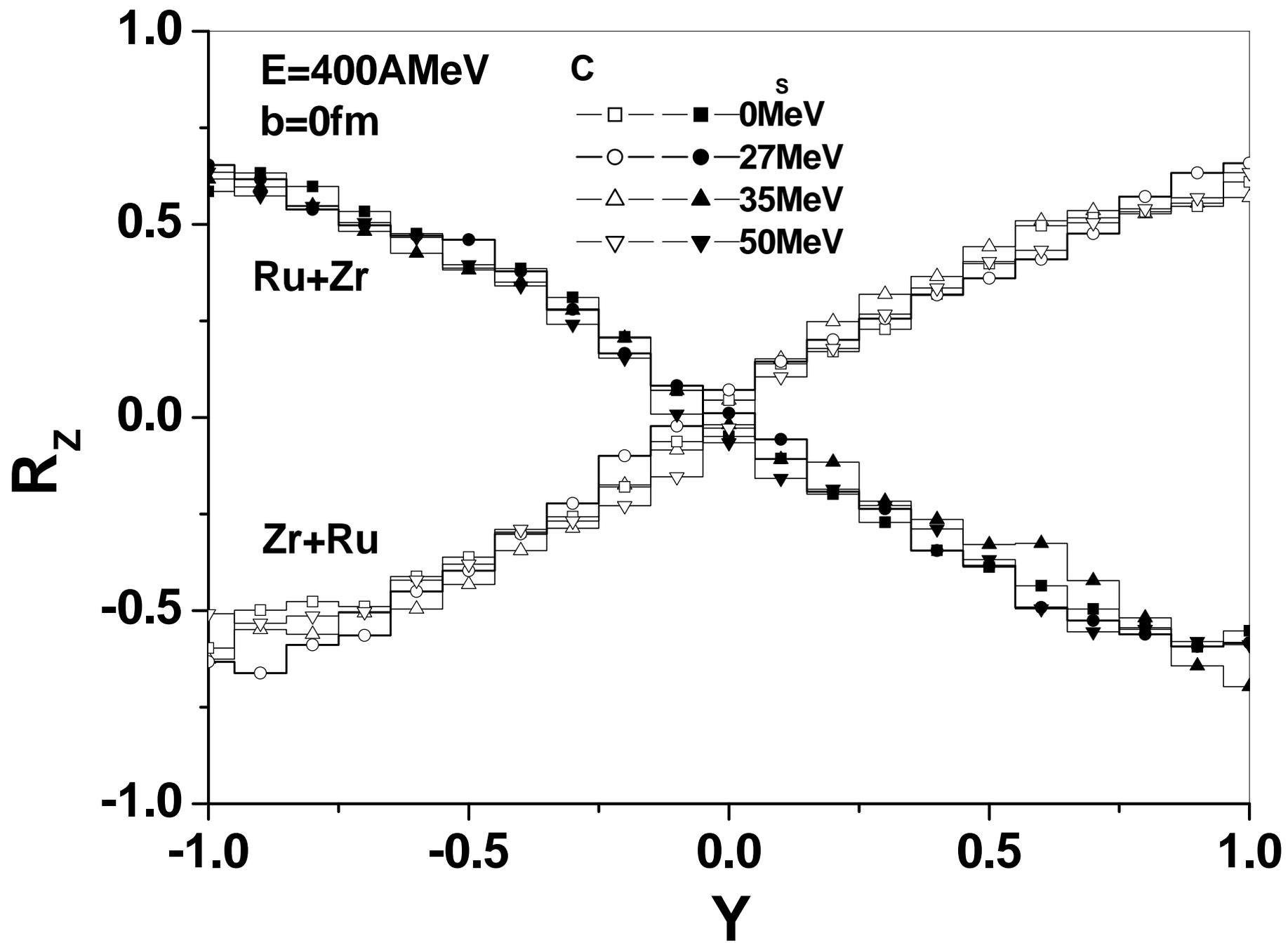
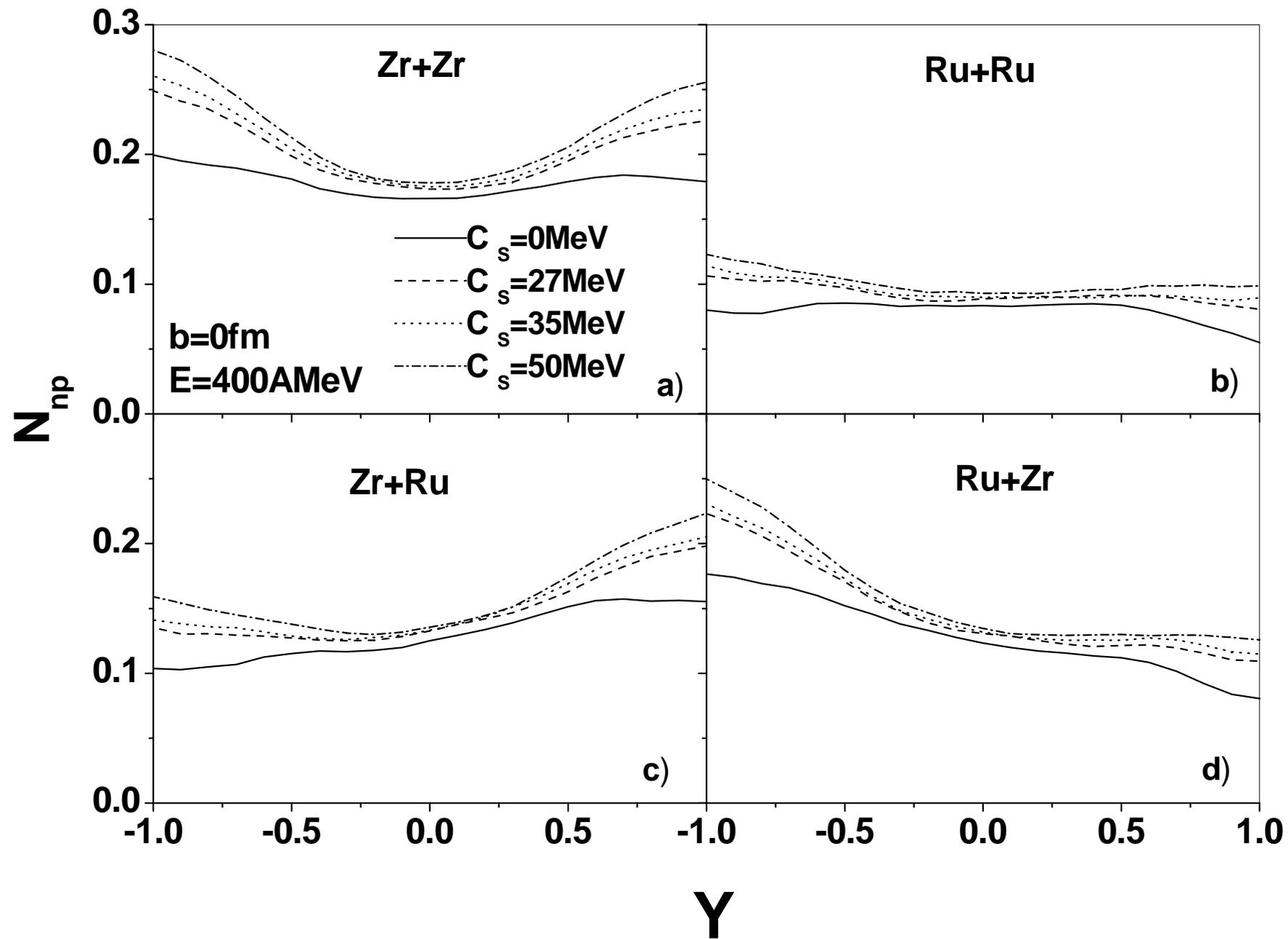
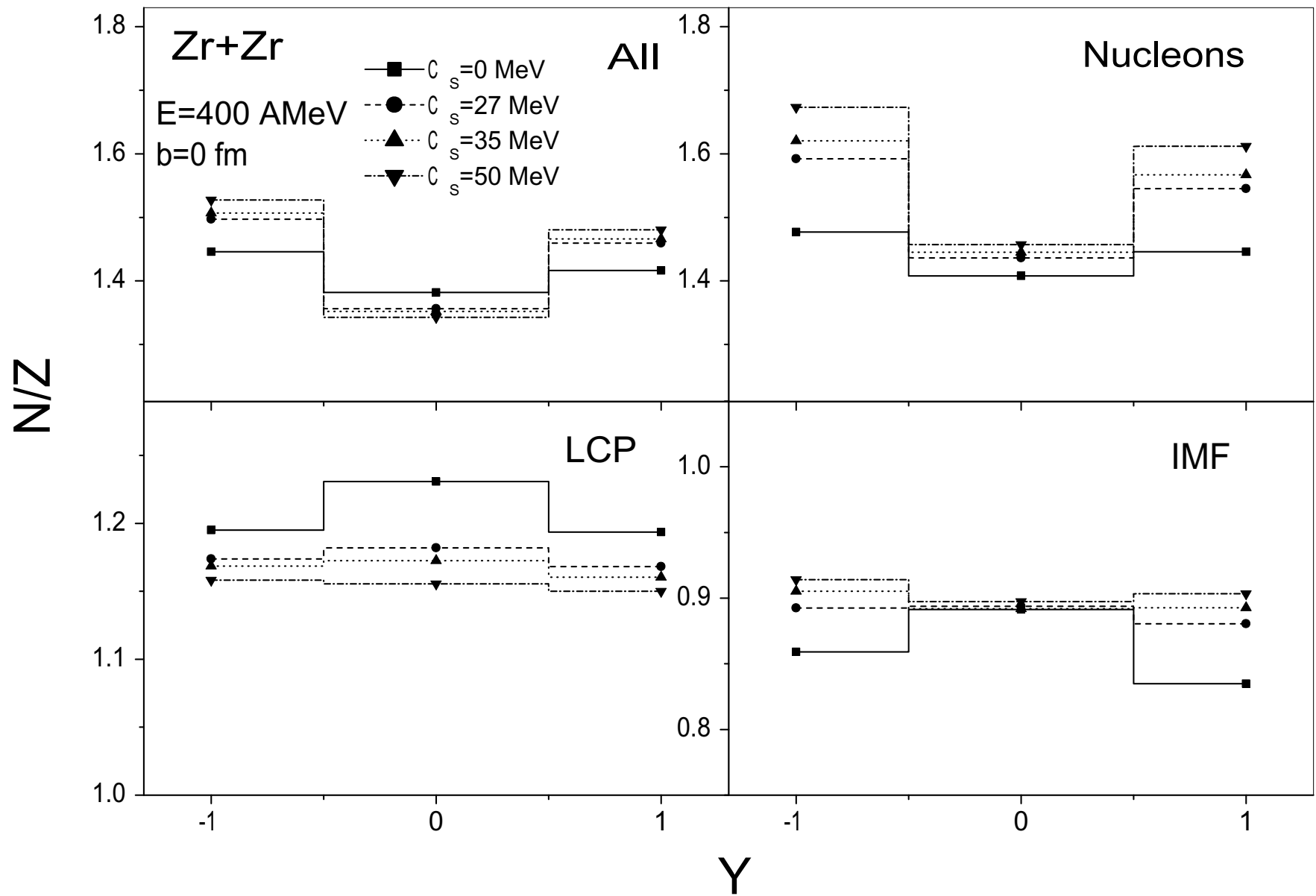
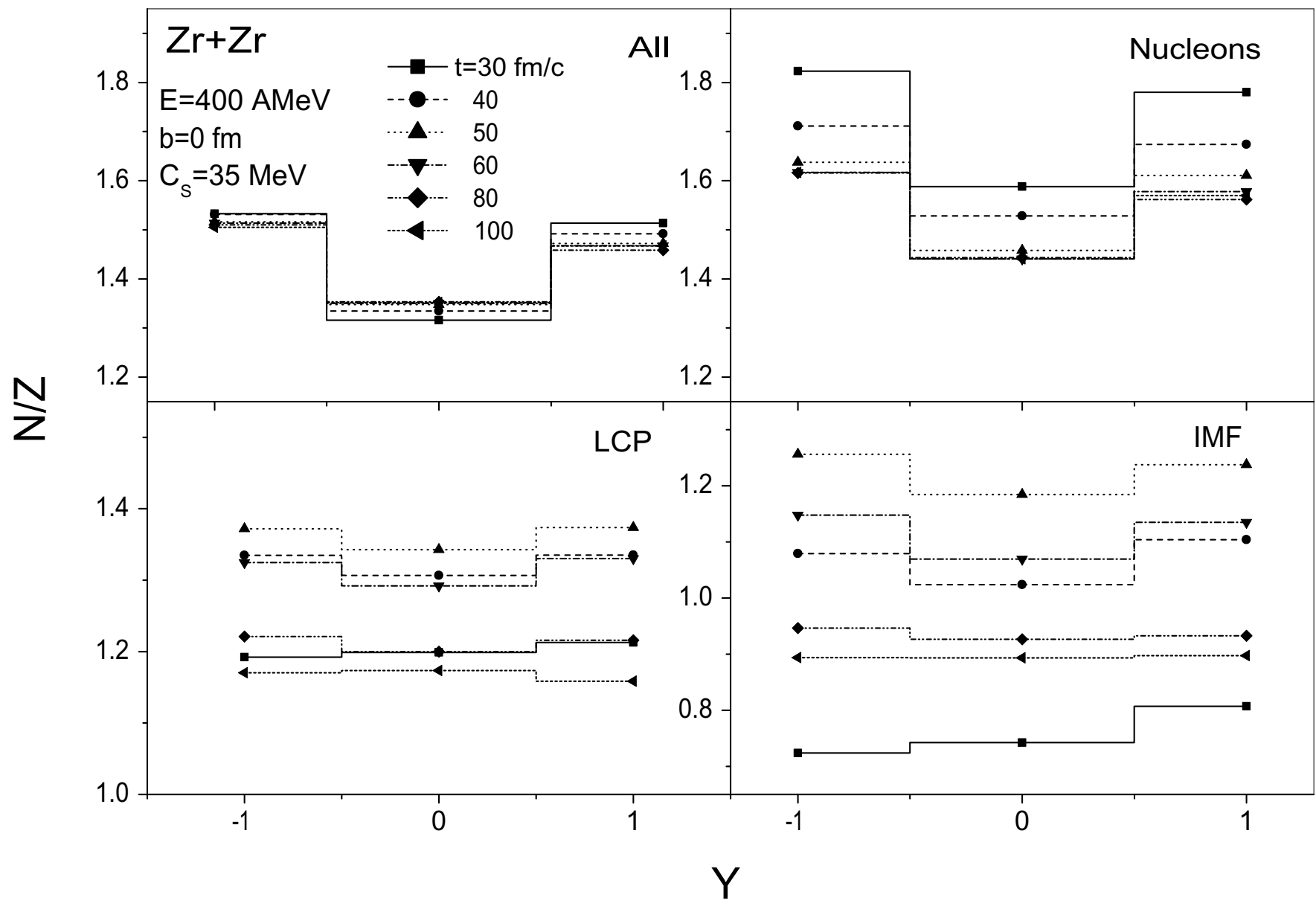
\end{document}